\documentclass[11pt]{article}
\usepackage{moriond,epsfig}

\newcommand{\epem}{\ensuremath{\mathrm{e}^+\mathrm{e}^-}}
\newcommand{\ppbar}{\ensuremath{\mathrm{p}{\overline\mathrm{p}}}}

\newcommand{\Zz}{\ensuremath{{\mathrm{Z}^0}}}
\newcommand{\WW}{\ensuremath{\mathrm{W}^+\mathrm{W}^-}}

\newcommand{\QQ}{\ensuremath{\mathrm{qq}}}
\newcommand{\ff}{\ensuremath{{f\overline{f}}}}
\newcommand{\lnu}{\ensuremath{\ell\nu_\ell}}

\newcommand{\tnu}{\ensuremath{\tau{\nu}_{\tau}}}

\newcommand{\taunu}{\ensuremath{\overline{\nu}_{\tau}}}

\newcommand{\qqqq}{\ensuremath{\QQ\QQ}}
\newcommand{\qqln}{\ensuremath{\QQ\lnu}}

\newcommand{\lnln}{\ensuremath{\lnu\lnu}}

\newcommand{\Mz}{\ensuremath{M_{\mathrm{Z}}}}
\newcommand{\Mw}{\ensuremath{M_{\mathrm{W}}}}

\newcommand{\Gw}{\ensuremath{\Gamma_{\mathrm{W}}}}
\newcommand{\Gz}{\ensuremath{\Gamma_{\mathrm{Z}}}}

\newcommand{\OPAL}{\mbox{OPAL}}
\newcommand{\ALEPH}{\mbox{ALEPH}}
\newcommand{\DELPHI}{\mbox{DELPHI}}
\newcommand{\LT}{\mbox{L3}}

\newcommand{\LEP}{\mbox{LEP}}

\newcommand{\roots}{\ensuremath{\sqrt{s}}}

\newcommand{\Zgamma}{\ensuremath{\Zz/\gamma}}
\newcommand {\ee}         {\ensuremath{\mathrm{e}^+ \mathrm{e}^-}}

\def\etal{\mbox{{\it et al.}}}

\def\gappeq{\ensuremath{\mathrel{ \rlap{\raise.5ex\hbox{>}}
                      {\lower.5ex\hbox{\sim}}}}}
\def\lappeq{\ensuremath{\mathrel{ \rlap{\raise.5ex\hbox{<}}
                      {\lower.5ex\hbox{\sim}}}}}

\newcommand{\PLB}[3]  {Phys.\ Lett.\ \textbf{B#1} (#2) #3}
\newcommand{\ZPC}[3]  {Z.\ Phys.\ \textbf{C#1} (#2) #3}
\newcommand{\EPC}[3]  {Eur.\ Phys.\ J.\ \textbf{C#1} (#2) #3}

\newcommand{\PRL}[3]  {Phys.\ Rev.\ Lett.\ \textbf{#1} (#2) #3}

\def\opalabbiendi{OPAL Collaboration, G.\ Abbiendi \etal}

\def\be{\begin{equation}}
\def\ee{\end{equation}}
\def\bea{\begin{eqnarray}}
\def\eea{\end{eqnarray}}

\begin{document}

\vspace*{4cm}

\title{STANDARD MODEL ELECTROWEAK MEASUREMENTS AT LEP}

\author{ FRANCESCO SPAN\`O }

\address{CERN, European Center for Nuclear Research\\
1211 Gen\`eve 23 - Switzerland\\
francesco.spano@cern.ch}

\maketitle\abstracts{
The current status of electroweak physics results from LEP is
reviewed. Particular emphasis is placed on the latest results on the
properties of the Z and W bosons. The updated status of the global
electroweak fit to the standard model and the resulting standard model
Higgs mass limits are presented.}

\vspace{-0.8cm}
\section{Introduction}
\vspace{-0.3cm}
Electron-positron collision allow a detailed investigation of the
standard model (SM) thanks to their clean experimental
conditions. 

The Large Electron-Positron collider (\LEP) 
was operated at CERN for about 12 years (1989 - 2000) at increasing
center of mass energies (\roots): both at the \Zz\ pole (\roots\ $\approx$
91 GeV) and beyond (\roots\ = 160-209 GeV). 
Its maximum  instantaneous luminosity was 0.5-1 $10^{32}$
$\mathrm{cm}^{-2}$ $\mathrm{s}^{-1}$ with a 45 kHz bunch-crossing rate.

The four \LEP\ multi-purpose detectors (\ALEPH, \DELPHI, \LT, \OPAL) 
collected an integrated luminosity of about 1000 $\mathrm{pb}^{-1}$ each (with
about 700 $\mathrm{pb}^{-1}$ beyond the \Zz\ pole ). These data correspond to
about 4.5 million \Zz\ and about 11000 W pair events per experiment.
\vspace{-0.4cm}
\section{\Zz\ physics}
\vspace{-0.2cm}
The \Zz\ boson couples to fermion-anti-fermion pairs. 
A gauge-invariant description embeds its production in 
the two-fermion production process \epem $\rightarrow$ \ff. 

They complement cross sections information to obtain the absolute size
of the parity-violating couplings. In addition they allow universality
tests by comparing $\sin\theta_{eff}^f$ and the $\rho_f$ parameter
($\rho_f$ =\Mw/(\Mz $\cos^{2}\theta_{eff}^f$) i.e. the ratio of
neutral and charged current interaction strengths).
\vspace{-0.4cm}
\subsection{Physics at the \Zz\ resonance}
\vspace{-0.2cm}
The \LEP\ \Zz\ pole results are final~\cite{theZ0}. 

\noindent $\bullet$ The two-fermion production cross section as a function of \roots\
yields both the \Zz\ mass, \Mz\ and its total width, \Gz . 
The ratios of cross sections for different two-fermion processes provide the
\Zz\ partial widths and information about the relative strength of the \Zz\
couplings to different final state fermions (see sections 1.4 and 1.5.1 of~\cite{theZ0}).

\begin{figure}[h]
\centerline{\psfig{figure=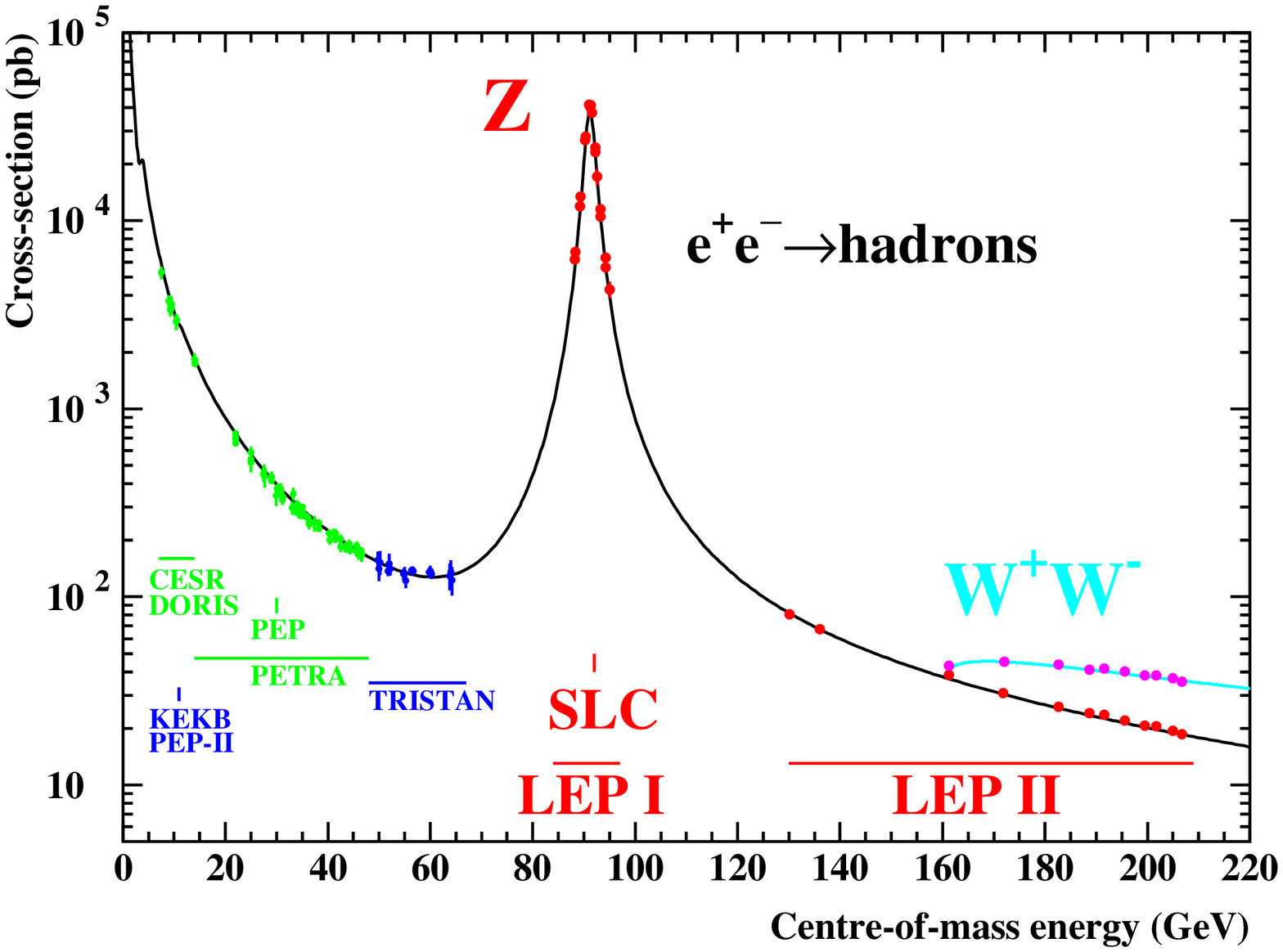,height=5cm} \psfig{figure=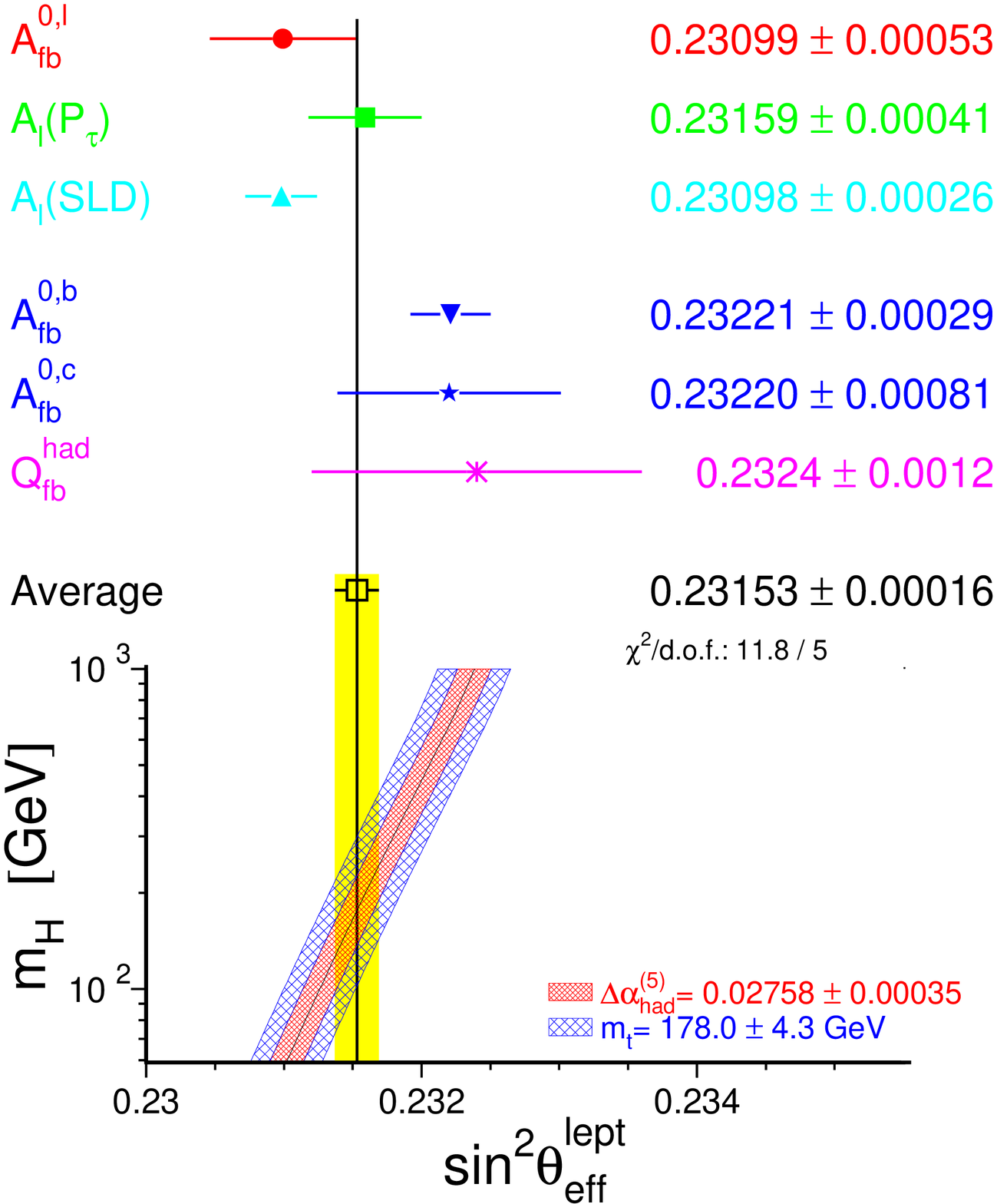,height=7cm}}
\caption{Left plot: \Zz\ hadronic cross section; points are
  measurements and the solid line is the SM prediction. Various \epem\
  collider energy ranges are reported.  
  Right plot: Comparison of the effective electroweak mixing angle
  $\sin\theta_{eff}^{lept}$ for leptons obtained
  from asymmetries depending on lepton couplings (top) only and also on quark
  couplings (bottom). The SM prediction is shown as a function of
  the Higgs mass. Vacuum polarization corrections and top mass
  uncertainty dominate the SM prediction uncertainty.
 \label{fig:zlineshape}}
\end{figure}

The \Zz\ lineshape parameters are known at the sub-per mil
level as shown in table~\ref{tab:z0tab}. A lineshape example is shown in
Figure~\ref{fig:zlineshape} (left). 
\begin{table}[ht]
\begin{center}
\begin{tabular}{|c|c||c|c|}
\hline
\Mz & 91.1875 $\pm$ 0.0021 GeV & $ \sigma^0_{had}$ & 41.540 $\pm$ 0.037 nb \\
 \Gz & 2.4952 $\pm$ 0.0023 GeV  & $\rho_l$ & 1.0050 $\pm$ 0.0010  \\ \hline

\end{tabular}
\caption{\Zz\ lineshape basic parameters (from
  top to bottom): the \Zz\ mass and  width, the hadronic cross section
  at the pole and the ratio of neutral to charged current interaction
  strength for leptonic final states at the pole. The extracted values
  have non zero correlation matrix.\label{tab:z0tab}}
\end{center}
\end{table}

A fundamental by-product~\cite{theZ0} is the determination of the
number of light neutrino families, $N_{\nu}$ = 2.9840$\pm$ 0.0082, by
comparing the \Zz\ invisible branching fraction measured at the pole 
($(\Gamma_{inv}/\Gamma_{ll})^{0}$ = 5.943 $\pm$ 0.016) with the
expectation obtained from the standard model \Zz\ decay into
neutrinos ($(\Gamma_{\nu\nu}/\Gamma_{ll})^{SM}$ = 1.99125 $\pm$ 0.00083). 

\noindent $\bullet$ The right and left-handed coupling of the \Zz\ to
fermions are different (see section 1.4 in~\cite{theZ0}) and
consequently violate parity invariance. In \epem\ collisions \Zz\
 bosons can then be expected to exhibit a net polarization along the
 axis of colliding beams, even when the incoming particles are not
 polarized.
 The decay of polarized \Zz's results in fermions with net helicity
 and with asymmetric angular distributions with respect to the beam
 direction. 

The effective electroweak mixing angle for
leptons, $\sin\theta_{eff}^{lept}$ is obtained at \LEP\ by measuring 
forward-backward asymmetries at the \Zz\ pole for different final
state fermions and using the corresponding left-right forward-backward
asymmetries measured at SLD to account for quark or electron
couplings (see section 1.5.3 of~\cite{theZ0}).
Three measurements are derived from leptonic final states
only: asymmetry from \Zz\ decay to leptons at the pole at \LEP\ is
complemented by $\tau$ polarization $P_\tau^{0}$
and left-right asymmetry $A_l$ at the SLD.
Quark final states at \LEP\ provide three complementary results using  
forward-backward asymmetries for b and c quarks final states at the
pole, ($A_{fb}^{0b}$, $A_{fb}^{0c}$) and the 
jet charge asymmetry. This is all summarized in
Figure~\ref{fig:zlineshape} (right). The combined result is
$\sin\theta_{eff}^{lept}$ = 0.23153 $\pm$ 0.00016.
\noindent $A_{l}$ and  $A_{fb}^{0b}$ show a 3.2 $\sigma$ discrepancy:
this is the main contributor to the 3.7$\%$  least
squared probability for the combined result. The systematic uncertainties in 
both measurements are considered to be under control (QCD and flavour
corrections for $A_{fb}^{0b}$, beam polarization for $A_{l}$) and the 
discrepancy is treated as a fluctuation~\cite{theZ0} rather than a
sign of new physics.
\vspace{-0.4cm}
\subsection{Two fermion physics above the $Z^{0}$ resonance }
\vspace{-0.2cm}
The cross sections and asymmetries in two fermion events are also
measured at higher \roots\ than the \Zz\ pole (\roots = 130 - 209
GeV). They test the standard model in a different physics regime: the
pure \Zz\ cross section decreases as the size of photon exchange and \Zgamma\
interference become important and the total cross section falls off.  
Measurements are performed both for inclusive and for high-energy non-radiative
events~\footnote{In addition to non-radiative events the inclusive
  sample consists of a sizeable fraction of radiative returns to the
  \Zz\ peak via initial state radiation~\cite{leprep}.}.
The LEP averages show good agreement with the standard model
predictions and help put limits on new physics that could lead to
visible effects in the selected final states at these energies, for
example from contact interactions or Z' bosons~\cite{lepsummary}.

\begin{figure}[htb]
\centerline{
\psfig{figure=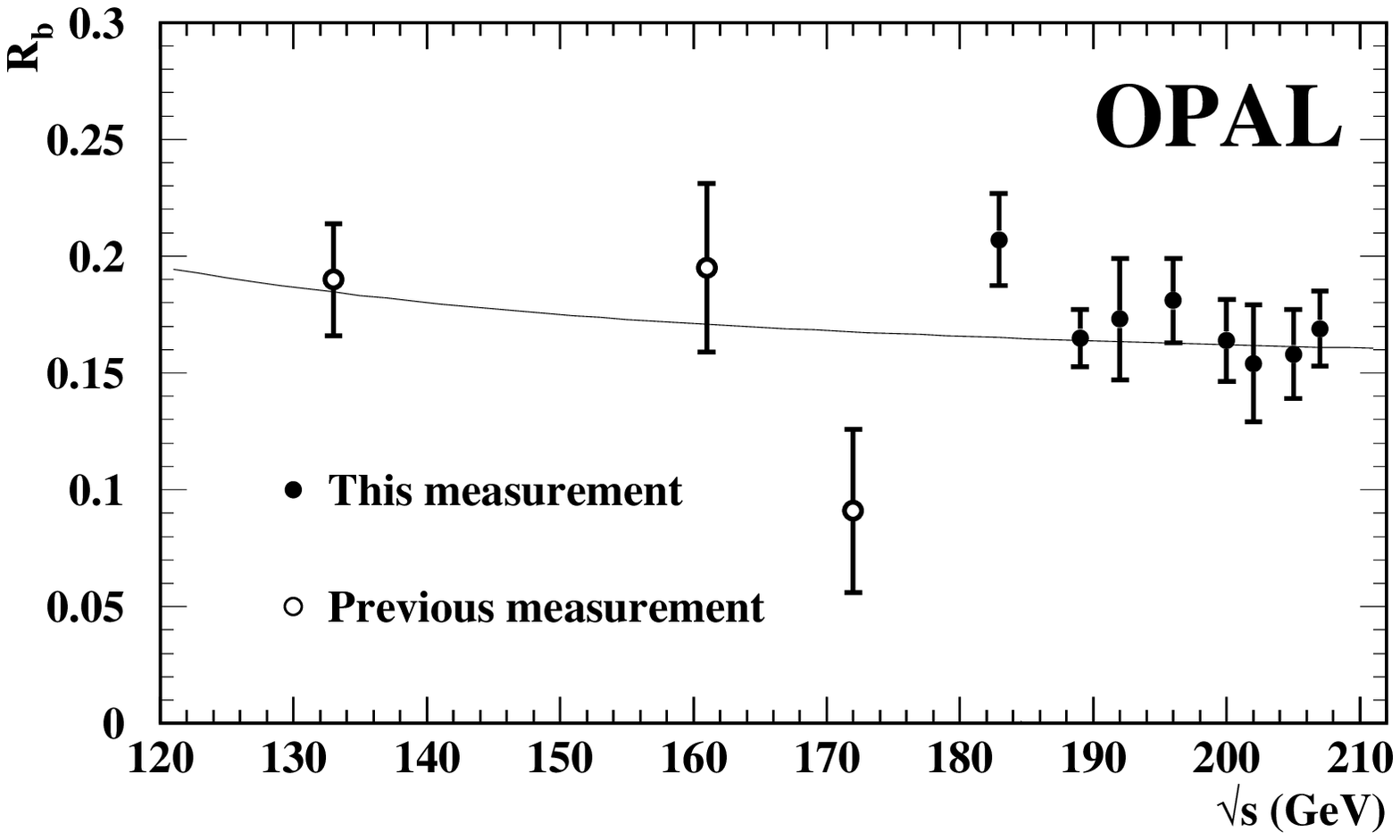,height=6cm} 
\psfig{figure=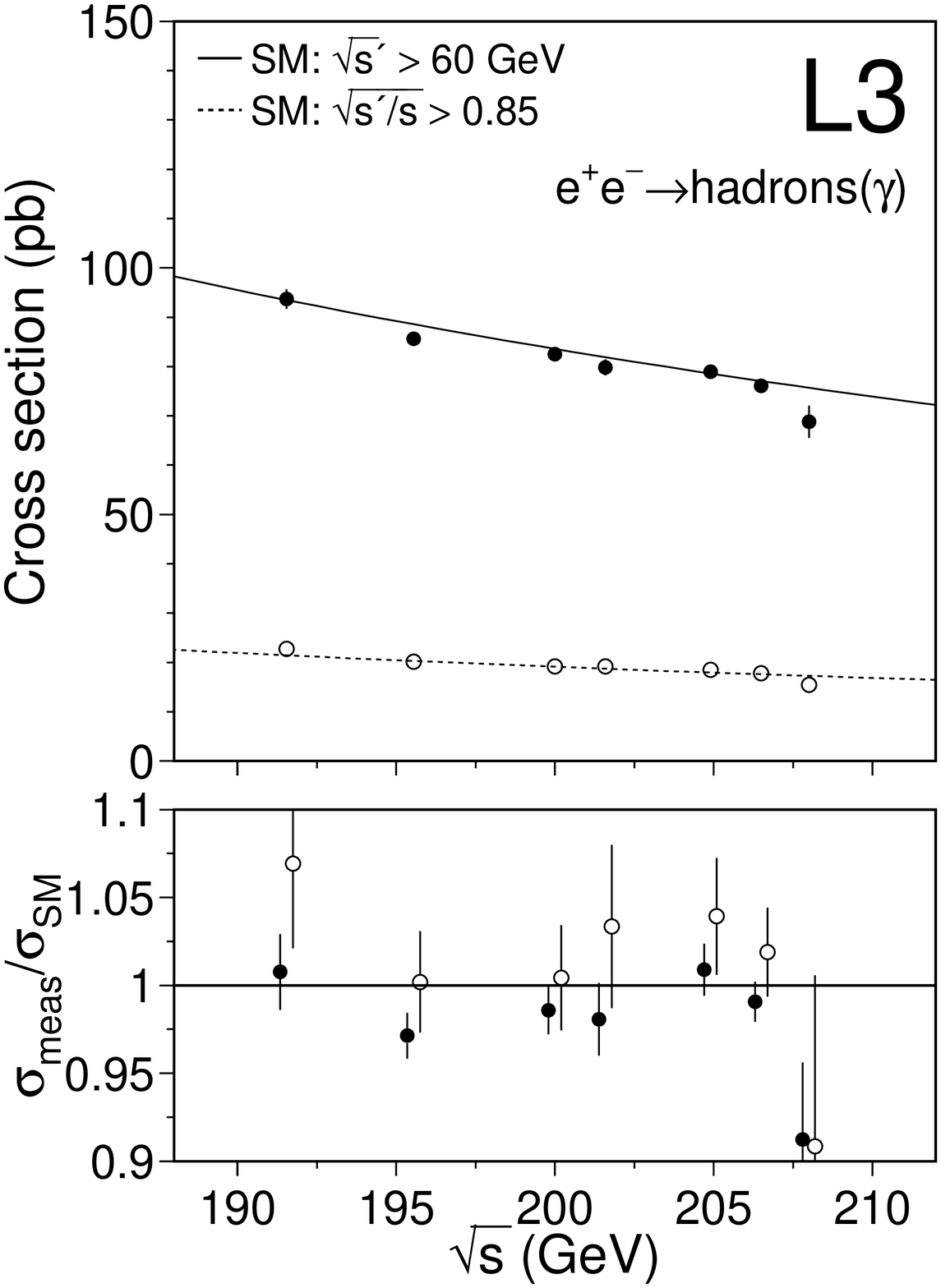,height=7cm}
}
\caption{Left plot: $R_b$ measurements from \OPAL\ (points with error
  bars) compared to the SM prediction (solid line). Right plots: cross section for
  hadrons($\gamma$) from \LT\ for the inclusive sample (filled symbols) and for
  the high energy sample (open symbols) compared to the SM predictions
  in solid (inclusive) and dashed (high-energy) lines (upper plot);
  ratios of measured cross sections to SM predictions (lower plot)}
\label{fig:2fbeyondz}
\end{figure}

Two of the most recent results are:

$\bullet$ \OPAL\ final measurement of $R_{\mathrm{b}}$~\cite{rbopal}(ratio of
the $\mathrm{b}\overline{\mathrm{b}}$ cross section to the
$\mathrm{q}\overline{\mathrm{q}}$ cross section in \epem collisions). $R_b$
values (in figure~\ref{fig:2fbeyondz} (left)) are compatible with SM
expectations. The mean ratio of eight $R_{\mathrm{b}}$ measurements to
their SM expectation is 1.055 $\pm$ 0.048.

$\bullet$ \LT\ final result~\cite{2ferml3} on hadron and lepton pairs cross
sections and lepton pairs asymmetries using both inclusive and non radiative
sample. An overall good agreement is found with the standard model.
An example is shown in figure~\ref{fig:2fbeyondz} (right) for hadron
cross section.
\vspace{-0.3cm}
\vspace{-0.2cm}
\section{W physics at LEP}
\vspace{-0.3cm}
W bosons are pair-produced at \LEP. The tree-level description of
\epem $\rightarrow$ \WW\ is the so-called CC03 diagrams~\cite{leprep}.
As each unstable W boson decays into lepton or quark pairs, a four
fermion (4f) final state is obtained with three possible topologies.
The fully leptonic channel \lnln\ is characterized by two high energy 
isolated acoplanar leptons with large missing energy. The
semi-leptonic channel \qqln\ exhibits an isolated high energy
lepton with two jets and missing energy. The \qqqq\ channel features
at least four jets and very little missing energy.
The three branching ratios are about 10\%, 46\% and 44\% respectively.
Width effects and interfering \epem $\rightarrow$ 4f diagrams destroy
CC03 gauge invariance. 
CC03 diagrams are embedded in an \epem $\rightarrow$ 4f
description~\cite{precisionworkshop} with O($\alpha$) electroweak
corrections that maintains gauge invariance and keeps 
theoretical uncertainties under control. This takes into account
background from non-WW \epem $\rightarrow$ 4f processes. The other
significant background is represented by \epem $\rightarrow$ \Zgamma
$\rightarrow$ hadrons.
\vspace{-0.4cm}
\subsection{W pair production}
\begin{figure}[ht]
\psfig{figure=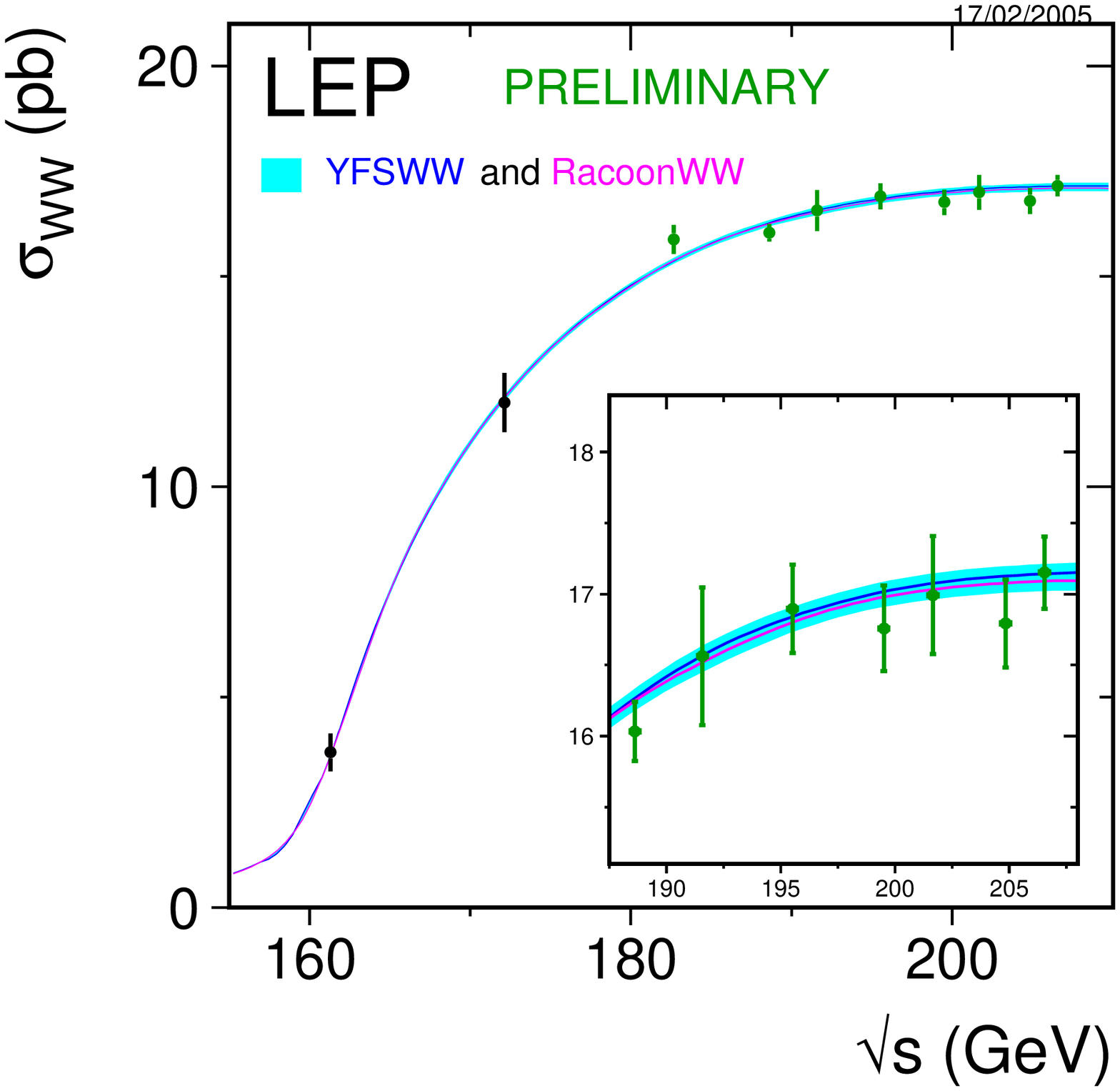,height=6cm}
\psfig{figure=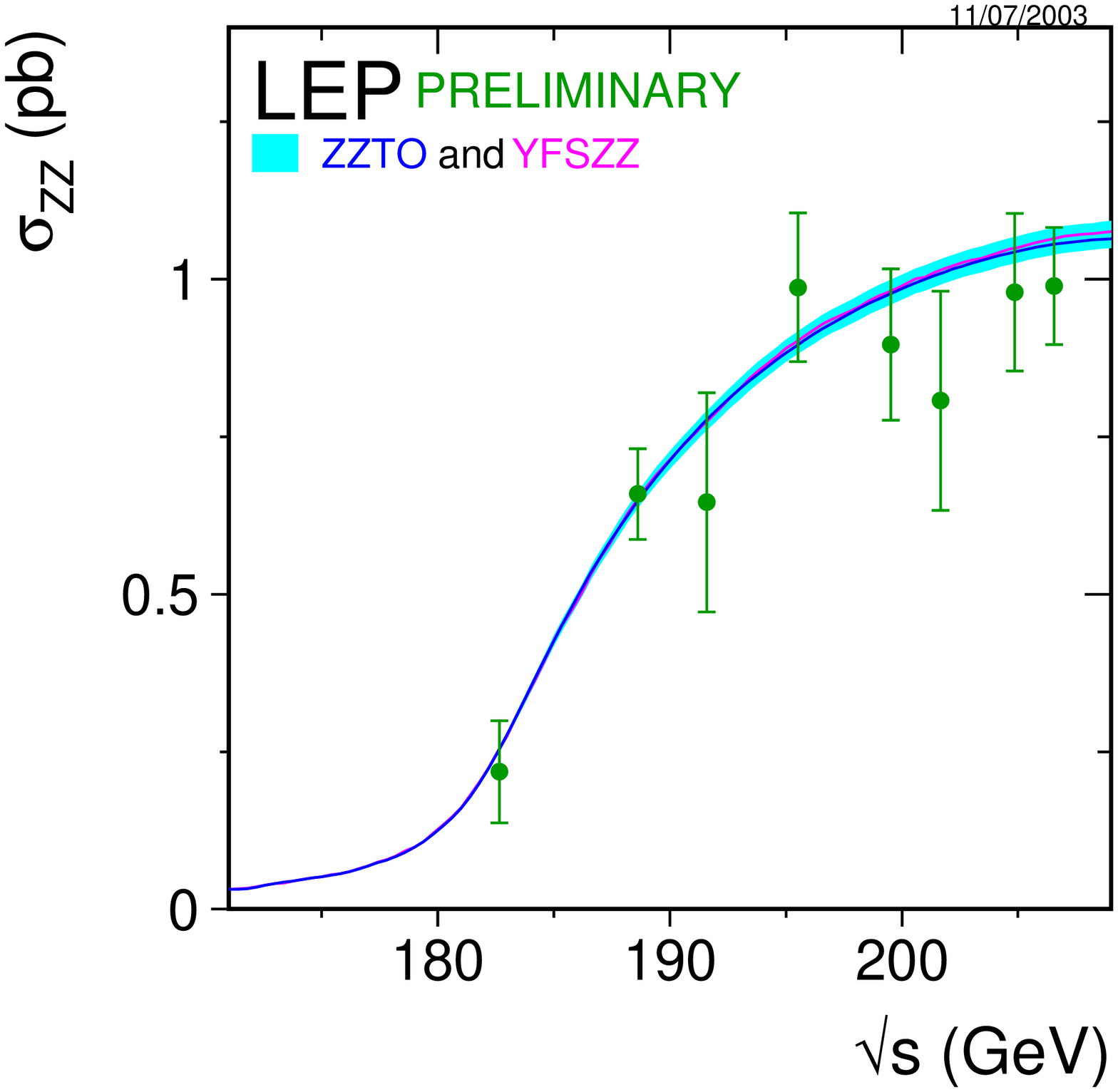,height=6cm}
\psfig{figure=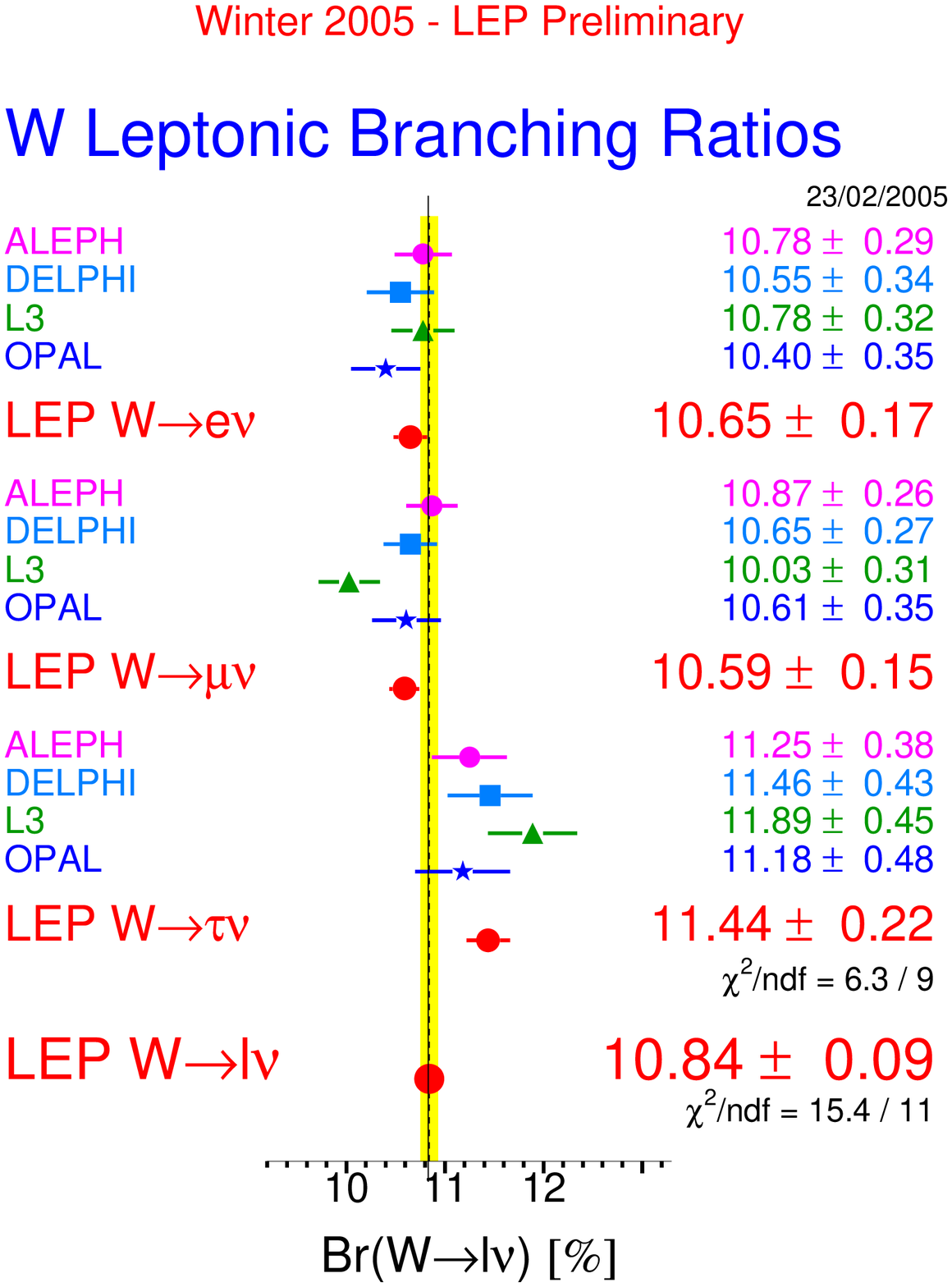,height=6cm}
\caption{\LEP\ combined results on W production and
  decay. Cross section
  measurement for W (left) and Z (center) pair production. W leptonic
  branching ratios are also shown (right).\label{fig:massshape}}
\end{figure}

\vspace{-0.2cm}
W pair production (CC03 cross section) in the kinematic
region explored by LEP shows a good consistency with the
SM expectations incorporating O($\alpha$) electroweak
corrections (only the W $\rightarrow$ \tnu branching ratio is
$\approx$ 2.8 $\sigma$ above its expectation). 
Final results for the \WW\ cross sections and W branching rations are
available from \ALEPH, \LT\ and \DELPHI. 
\OPAL\ has preliminary results for \roots\ = 161-
189 GeV and final for \roots\ = 192 - 207 GeV. 
Good agreement with the SM prediction is also found for \Zz\ pair
production (main 4f background to \WW\ after event selection). The
results~\cite{lepsummary} are shown in Figure~\ref{fig:massshape}. 
The typical cross section for \WW\ production beyond 180 GeV is about 17 pb. 
\vspace{-0.4cm}
\subsection{W mass and width extraction} 
\vspace{-0.2cm}
At threshold for \WW\ production (\roots\ $\approx$ 161 GeV), the  W
mass is derived from the cross section measurement. Above threshold,
real W bosons are reconstructed from their decay products and mass and
width extracted from appropriate event distributions.

Complex multi-step selections combining cut-based algorithms,
likelihood discriminants and neural networks are used to separate
signal from background.
Typical efficiencies are around 80\% for \qqln\ and \qqqq\ channels
and 70\% for \lnln\ channel. Typical purities are  85\% (\qqln),
80\% (\qqqq) and 90\% (\lnln).  

Above threshold \lnln\ events are not reconstructed due to the two
neutrinos in the final state and separate analyses~\cite{lepsummary}
are carried out.
To reconstruct the events, lepton identification in \qqln\ channel is 
carried out; the event remnant is forced into two jet.
Four jets are produced from \qqqq\ events.
DELPHI and OPAL allow for an additional gluon jet.
Event-by-event kinematic fit use knowledge of the precise beam energy
to constrain the total four-momentum.
The event-by-event mass resolution is greatly improved.
Various algorithms reduce jets-to-W mis-pairing in the 
\qqqq\ channel: consistency with W decay kinematics (\ALEPH, \OPAL),
multivariate selections and cuts in kinematic fit probability (\OPAL\
and \LT), combined information from all pairings (\DELPHI, \OPAL). The
resulting jet-to-W pairing efficiency is 70\%-90\%.

Reconstructed distributions are compared with the expectation by a
maximum likelihood technique to extract W mass and width.
Three different methods are used to estimate the expected data 
distributions as a function of \Mw\ and \Gw. 
An analytic asymmetric {\bf Breit-Wigner} function
(OPAL) provides a robust and simple method for preliminary
estimation. The analytic {\bf convolution} of a modified
Breit-Wigner~\footnote{To take into account initial
state radiation and phase space effects.} with 
an event-dependent detector response aims at using maximum information
to reduce statistical uncertainty (\DELPHI, \OPAL). 
Both analytic methods need Monte Carlo (MC) calibration to
correct biases. A {\bf re-weighting} technique (\ALEPH, \LT, \OPAL) generates 
expected data at arbitrary \Mw\ and \Gw\ using
a single fully simulated MC sample. This fully exploits the knowledge
encoded in the Monte-Carlo and minimizes possible bias effects.
\vspace{-0.4cm}
\subsection{W mass uncertainties}
\vspace{-0.2cm}
The LEP combined uncertainties on W mass as of Winter
2006~\cite{lepsummary} are illustrated in Table~\ref{tab:werr}.
The results are final for ALEPH~\cite{wmassaleph}, L3~\cite{wmassl3}
and OPAL~\cite{wmassopal}. DELPHI results are still preliminary.

\begin{table}[htbp]
\begin{center}
\caption{W mass uncertainties as of Winter 2006. \label{tab:werr}}
{\footnotesize
\begin{tabular}{@{}ccccc@{}} 
\hline
\multicolumn{2}{c}{Source} &\multicolumn{3}{c}{Uncertainties on \Mw (MeV)}\\[1ex]
\hline
{} & {} &{} &{} &{} \\[-1.5ex]
{} & {} & \qqln & \qqqq & Combined \\[1ex]
\multicolumn{2}{l}{QED (ISR/FSR,etc)}    &  9 &  5 & 8 \\[1ex]
\multicolumn{2}{l}{Hadronisation}        & 14 & 20 & 15 \\[1ex]
\multicolumn{2}{l}{Detector Systematics} & 11 & 8 & 10 \\[1ex]
\multicolumn{2}{l}{LEP beam energy}      & 9 & 9 & 9 \\[1ex]
\multicolumn{2}{l}{Colour reconnection} & - & 31 & 7 \\[1ex]
\multicolumn{2}{l}{Bose-Einstein Correlation} & - & 13 & 3 \\[1ex]
\multicolumn{2}{l}{Other} & 3 & 11 & 4 \\[1ex]
\hline
\multicolumn{2}{l}{Total Systematic} & 22 & 43 & 24 \\[1ex]
\multicolumn{2}{l}{Statistical} & 31 & 43 & 26 \\[1ex]
\hline \hline
\multicolumn{2}{l}{Overall} & 38 & 61 & 35 \\[1ex]
\hline
\end{tabular}\label{tableerr}}
\vspace*{-13pt}
\end{center}
\end{table}

The main updates are as follows:

{\bf LEP beam energy} The final \LEP\ energy
calibration~\cite{lepebeam} helps decrease the associated uncertainty
on \Mw.

{\bf Bose-Einstein Correlation (BEC)} \hspace{0.2cm} Unaccounted
quantum interference of identical bosons during
 hadronisation can correlate pions from different W bosons (inter-W
 BEC) and bias the \Mw\ and \Gw\ measurements.
 The uncertainties are set using the LUBOEI~\cite{bec}
 model taking the difference between the presence and
 the absence of such correlations.
 Current studies at LEP show no evidence of inter-W correlations 
 {\em \`a la} LUBOEI and the `percentage' of LUBOEI Inter-W
 correlation present in data is shown to be linear in \Mw\
 uncertainty~\cite{lepsummary}. This is used, for each experiment, to
 achieve a data-driven reduction of \Mw\ uncertainty (by 30\%) used
 for the combined result.
 
{\bf  Colour Reconnection (CR)} \hspace{0.2cm}
  Non-simulated colour cross-talk between decay products of
  different W bosons in \qqqq\ channel can bias the \Mw\ and \Gw\ determination.
  Different models predict mass biases up to 200 MeV. 
  The largest bias is foreseen by the SK model~\cite{cr}
  with variable CR strength {\em k}.
  LEP experiments use a `particle flow' technique~\cite{lepsummary}  to
  establish a 68\% C.L. limit on {\em k}. This provides a data-driven 
  uncertainty on \Mw\ and \Gw.  
  The major improvement is obtained by using estimates of jet
  angles which have low sensitivity CR effects: cutting out low
  momentum particles, reconstructing jets with cones of variable size,
  weighting momenta with a power of their magnitude.
  All final result use a momentum cut technique: with respect to
  previous measurements, a 15\% to 35\% increase in
  \Mw\ statistical uncertainty and a 30\%-100\% increase in
  \Mw\ hadronization uncertainty are more than compensated by a two to
  three-fold reduction in \Mw\ CR uncertainty.

  The combined \qqqq\ \Mw\ uncertainty  improves by  without biasing
  the W mass result. The \qqqq\ channel reaches a 23\% weight in the
  combineation (up from 9\% before BEC and CR
  reductions)~\footnote{\Mw\ statistical uncertainty without including
  final state interactions effects would be 21 MeV, the current 26 MeV
  value shows that most of \qqqq\ statistical power is being used.}. 

Other uncertainties incorporate ignorance on 
photon radiation in \epem $\rightarrow$ 4f,  MC statistics and
experiment-specific effects. 
\vspace{-0.35cm}
\subsection{LEP and global combined results for W mass: status and outlook}
\begin{figure}[ht]
\centerline{
\psfig{figure=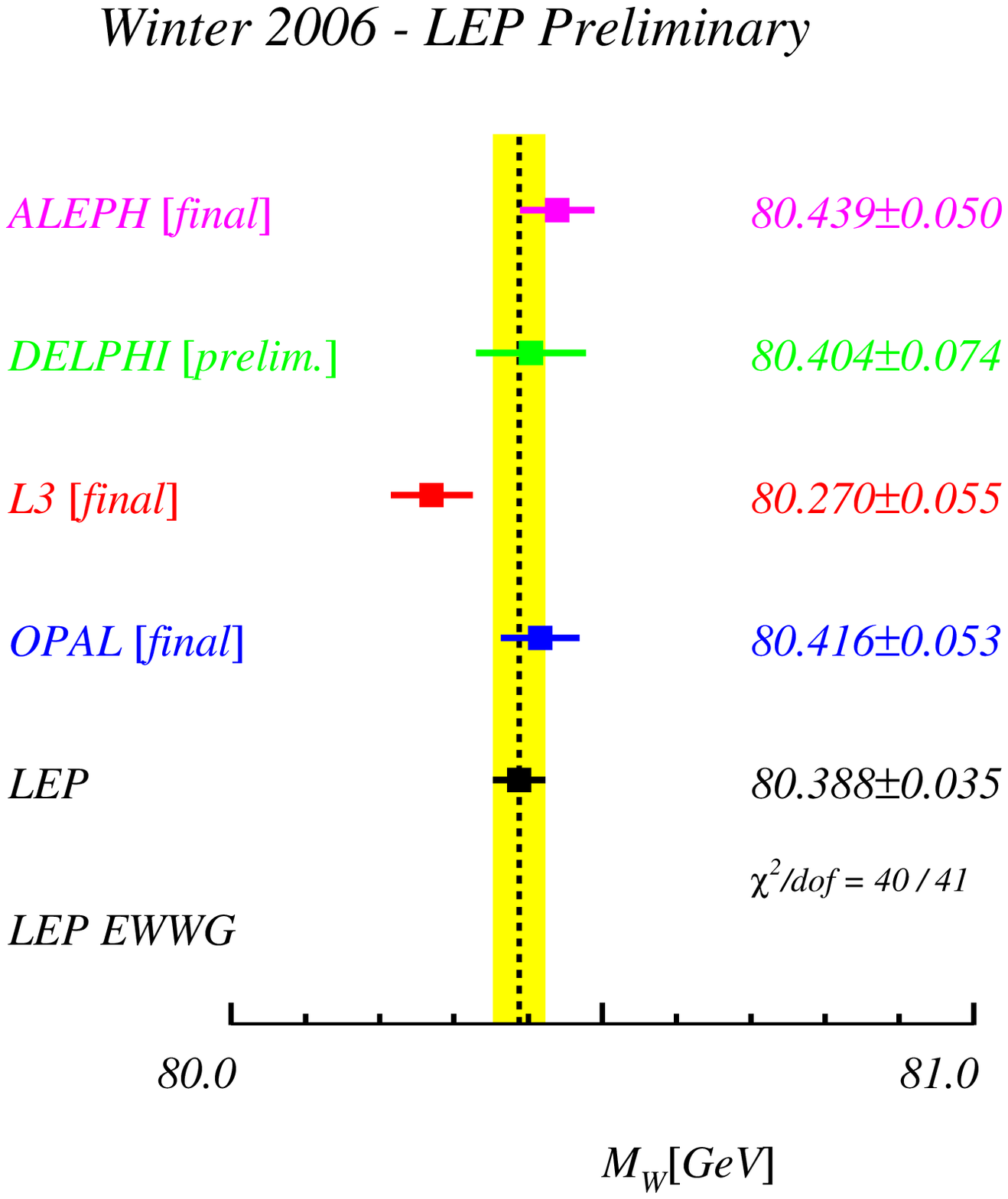,height=7cm}
\psfig{figure=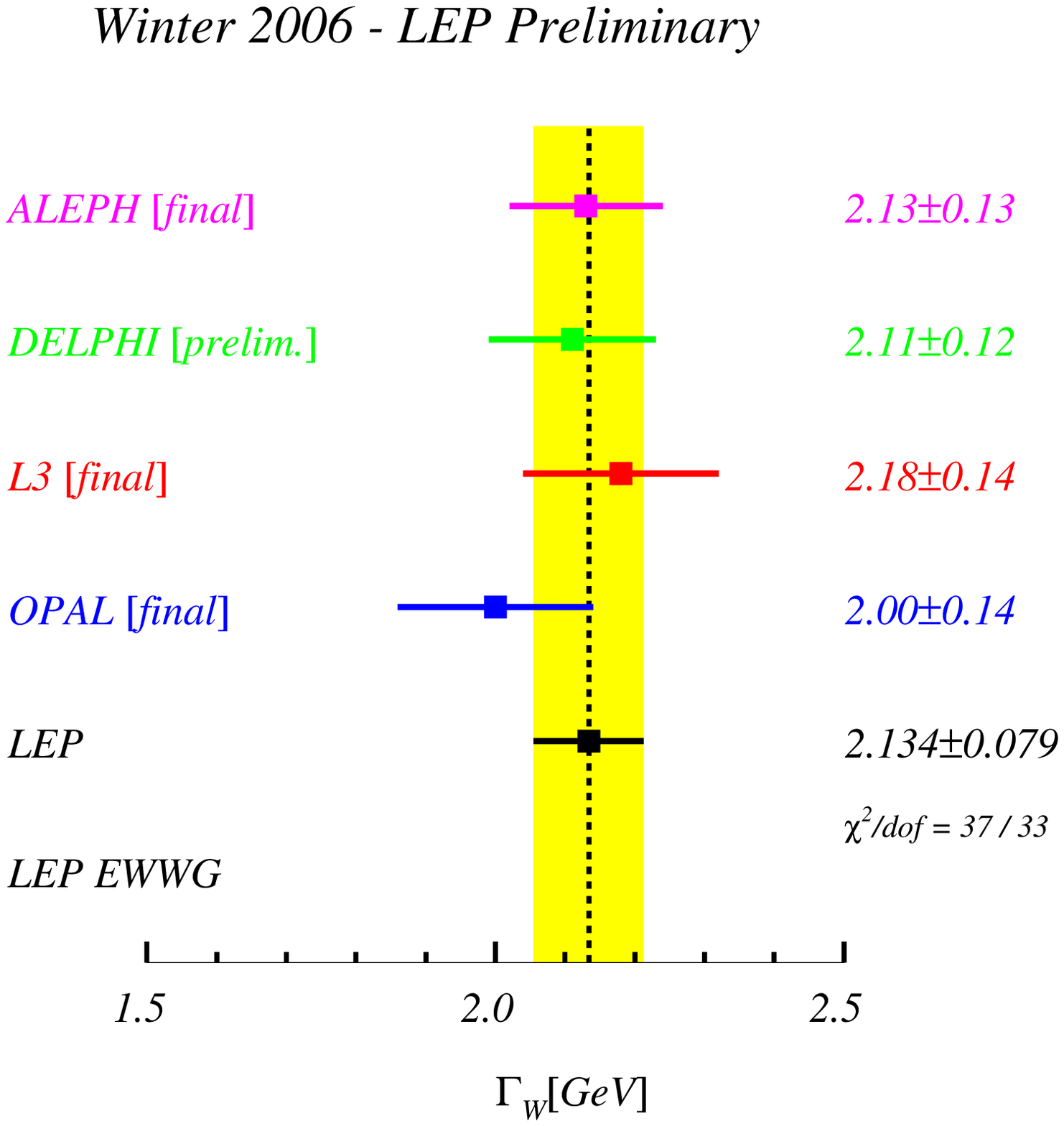,height=7cm}}
\caption{LEP W mass and width measurement as of Winter 2006}
\label{fig:lep2comb}
\end{figure}

\vspace{-0.2cm}
The latest combined results~\cite{lepsummary} for W mass and width
(Winter 2006) are shown in Figure~\ref{fig:lep2comb}.

The theoretical prediction for \Mw\ derived from the SM and the 
electroweak measurements is in good agreement with the measured LEP value. 
Results from \WW\ threshold are included in the LEP combination. 
The updated LEP results are \Mw\ = 80.388 $\pm$ 0.035 GeV and \Gw\ = 2.134
$\pm$ 0.079 GeV.

\begin{figure}[h]
\centerline{
\psfig{figure=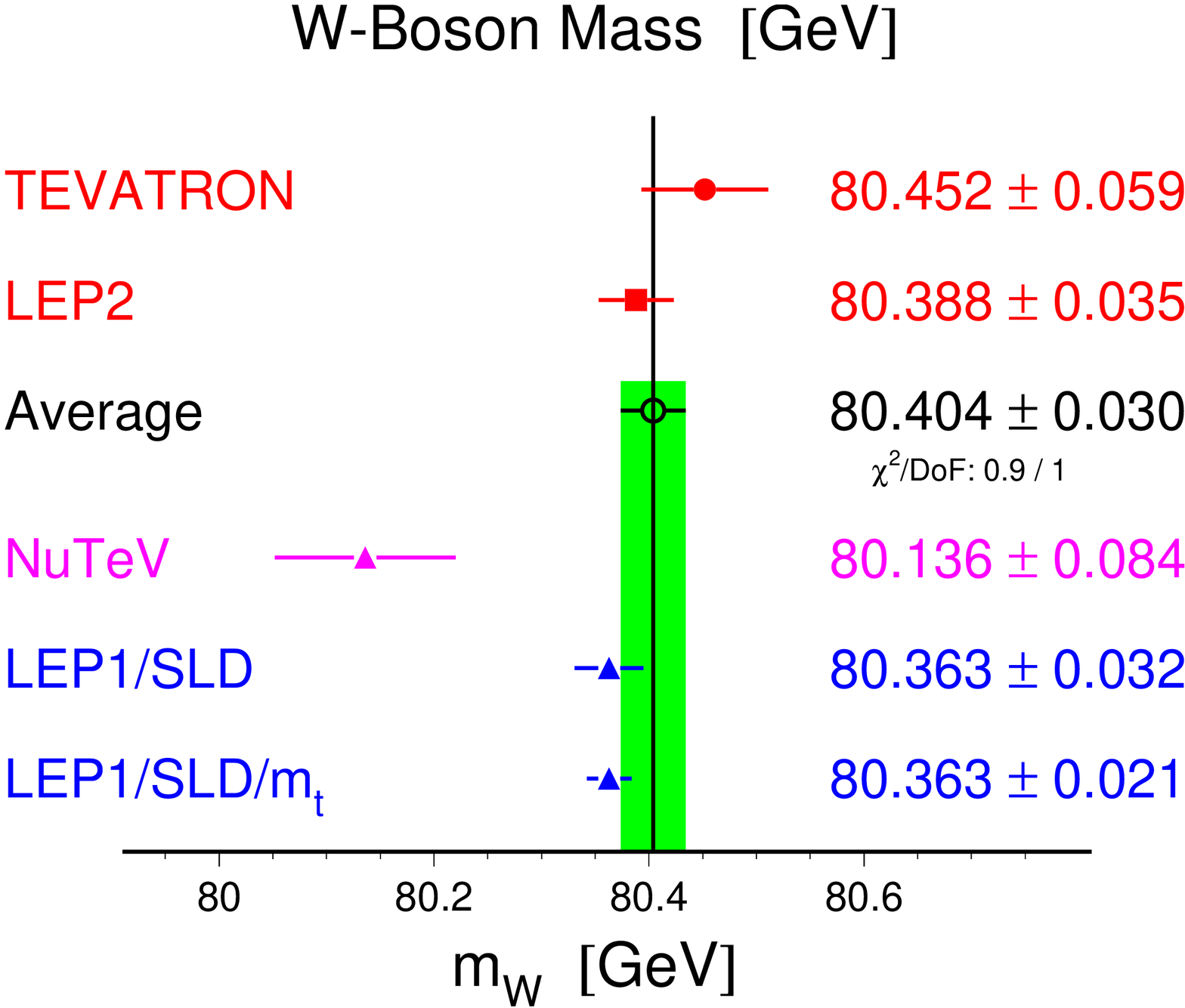,height=5cm}
\psfig{figure=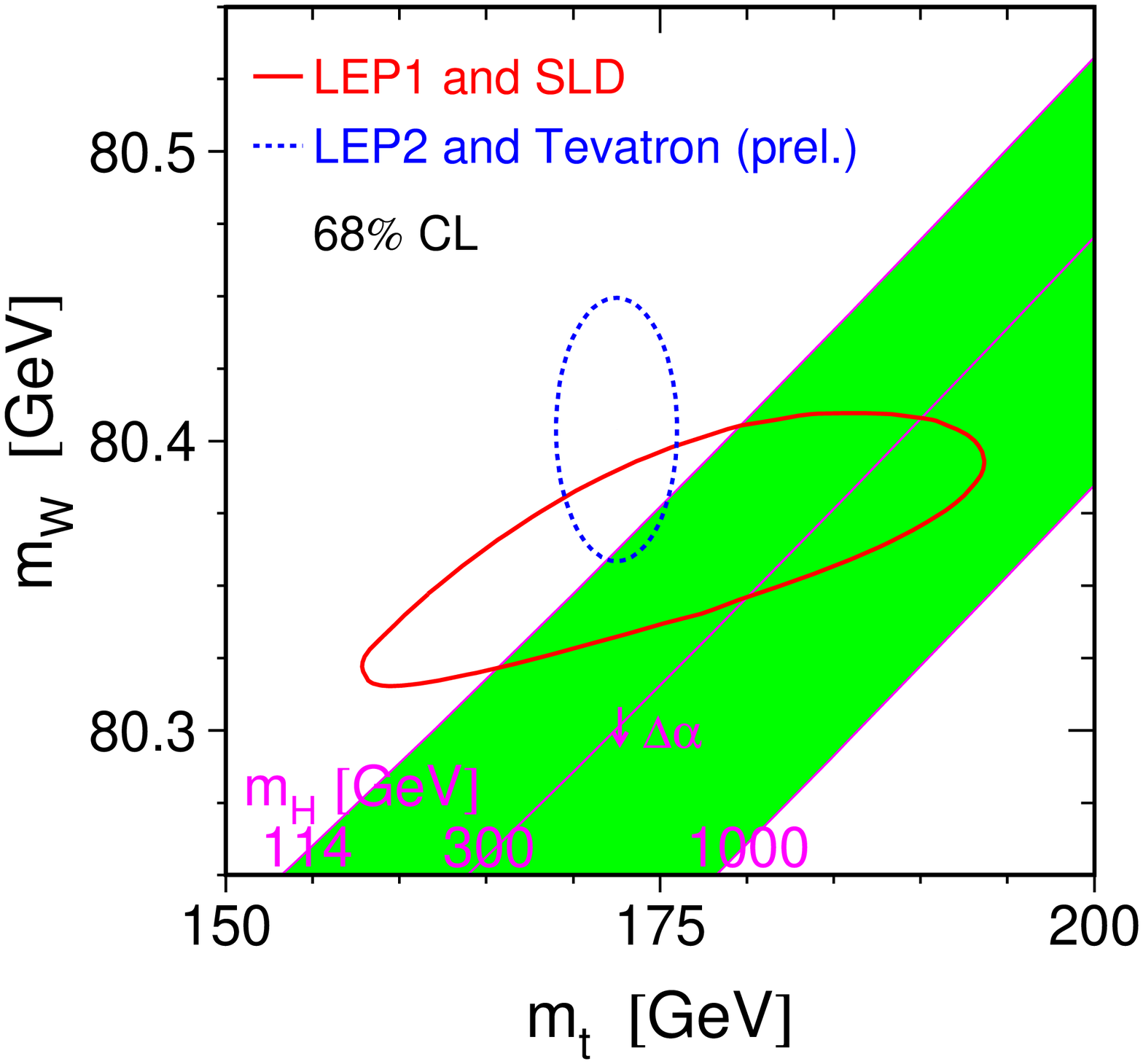,height=5cm}
}
\caption{Direct and indirect W mass measurements (left) and SM
  consistency plot (right)}
\label{fig:mwsummary}
\end{figure}

The direct measurement of the W mass at LEP and at \ppbar\ colliders
(CERN SppS, Tevatron) are in very good agreement. 
The indirect results do not show any significant discrepancy with the 
possible exception of the NuTeV value.
SM predictions based on the  W mass and Top quark mass prefer a
`low' (below $\approx$ 219 GeV) SM Higgs mass value.
The updated results~\cite{ewkwg} are summarized in Figure~\ref{fig:mwsummary}.

The final \LEP\ result for W mass and width will profit from the
final \DELPHI\ result and will use combined \LEP\ results on final
state interaction parameters for a coherent reduction of the
uncertainties. A reasonable target for the final LEP W mass
uncertainty is an improvement of 1 or 2 MeV on the present value.
\vspace{-0.4cm}
\section{Standard model status: global fit and the Higgs}
\begin{figure}[h]
\centerline{
\psfig{figure=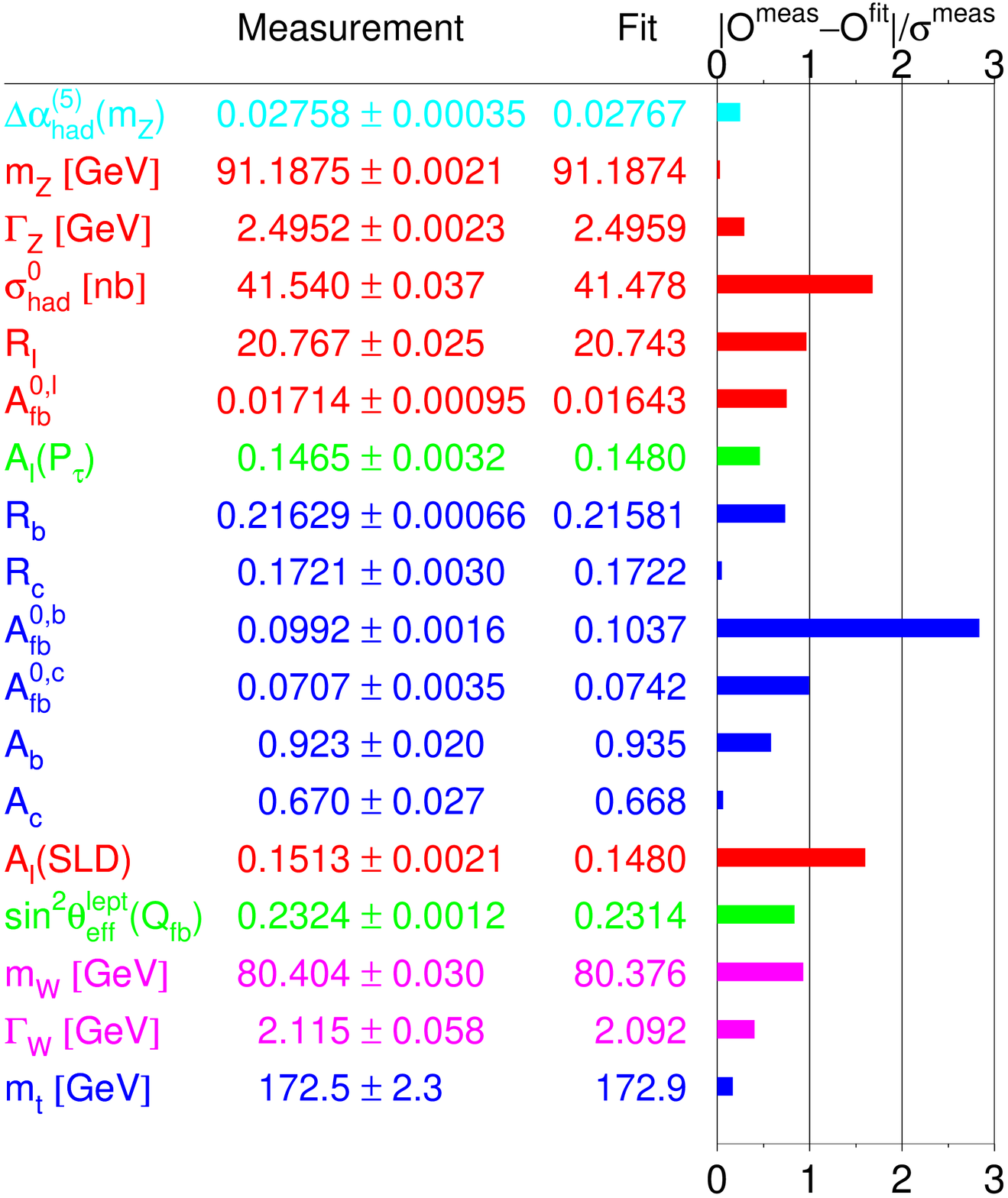,height=8cm}
\psfig{figure=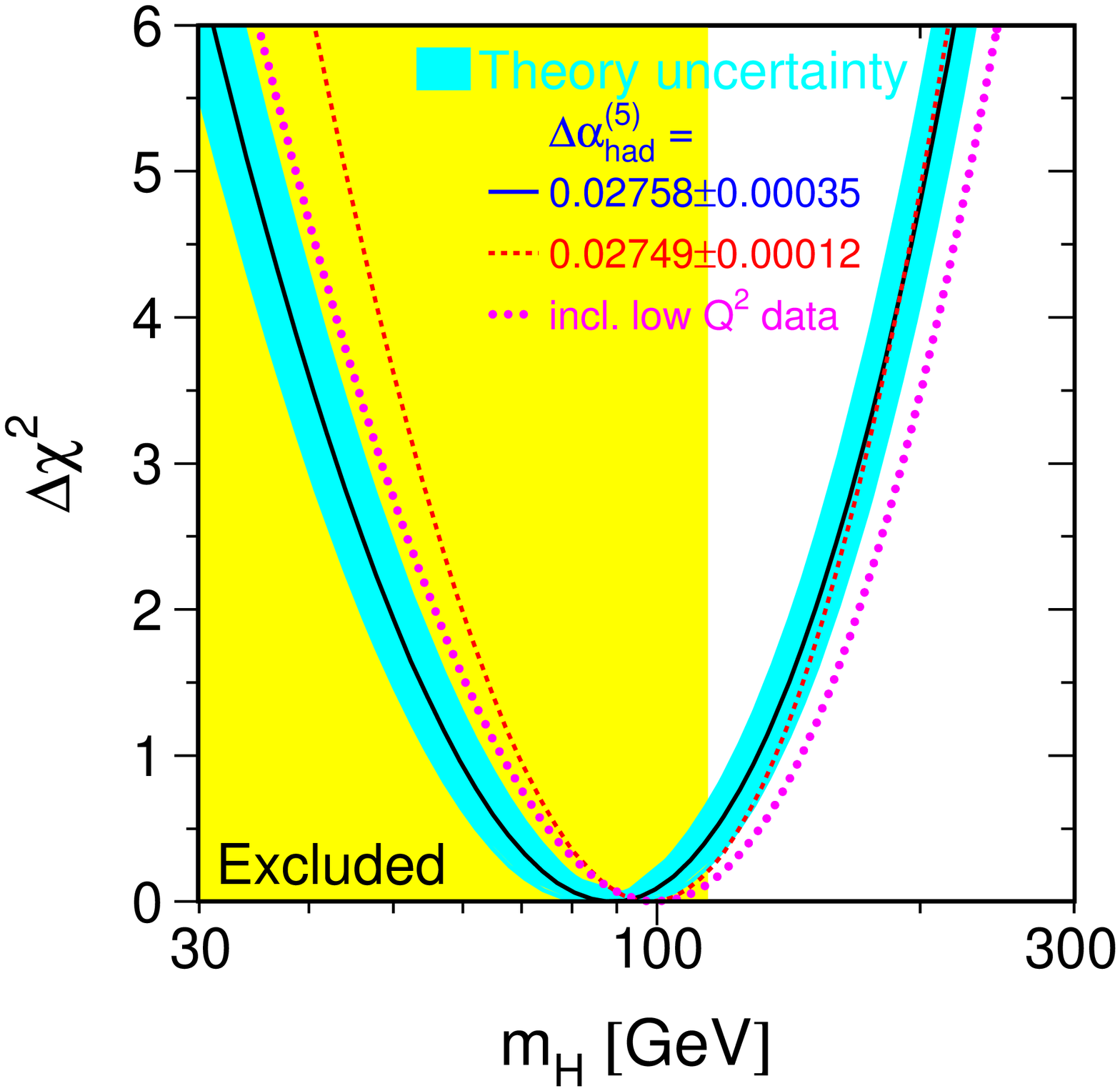,height=7cm}
}
\caption{Left: Comparison of the measurements with the SM expectation and
  the pulls they apply in the global fit. Right: $\delta
  \chi^{2}$($m_H$)= $\chi^{2}_{min}(m_{H})$ - $\chi^{2}_{min}$ of global
  SM fit as a function of $m_H$ compared to the 95\% yellow exclusion zone from \LEP\
   and including theoretical uncertainties (band), the effect of different
   estimates for $\alpha_{QED}$ renormalization at the \Zz\ pole and
   the impact of adding low $Q^{2}$ measurements in the fit. 
   The used top and W mass are the updated results for
   Winter 2006. \label{fig:smfit}}
\end{figure}

\vspace{-0.2cm}
A global least-squared fit is performed to extract five SM parameters
from which all the other oservables depend. The SM consistency
requires that all the observables are determined as a function of the
same free parameter values. The chosen parameters are: the QED and QCD
coupling constants at the Z pole, $\alpha_{QED}(\Mz)$,
$\alpha_{QCD}(\Mz)$, the masses of the Higgs boson, the top quark
and the Z boson.
Eighteen observables measured in high momentum transfer events
($Q^2$ $\approx$ $\Mz^{2}$) are used in the fit. The result is
shown in Figure~\ref{fig:smfit}(left). The fit $\chi^{2}$/d.o.f is
17.5/13 (probability($\chi^{2} > \chi^{2}_{min}$) = 18\%).
The largest contribution to the fit $\chi^2$ derives from
$A_{fb}^{0b}$ (about 2.8 $\sigma$) $A_{fb}^{0b}$ favours
high values of the Higgs mass, in contrast with \Mw\ and the leptonic 
asymmetries. The high $Q^2$ parameters are used to derive predictions for low
momentum transfer observables ($Q^2 \ll \Mz^{2}$). Good agreement is
observed for parity violation in atoms and in Moeller scattering.
The combination of left-handed effective coupling constants $g_{\nu Lud}$
derived from neutrino-nucleon scattering events in the NuteV
experiment shows a discrepancy of about 2.8 $\sigma$ with respect to
its SM expectation. A sizeable theoretical effort is ongoing on 
uncertainties of radiative corrections and QCD effects affecting the
measurement (see sec. 8.3.3 of~\cite{theZ0}).
\newline \indent The resulting limits on the standard model Higgs mass~\cite{lepsummary}
derived from the high $Q^2$ fit are shown in Figure~\ref{fig:smfit}
(right). At 95$\%$
confidence level, $M_H$ is below 186 GeV ($>$ 114.4 GeV from LEP
direct searches). If \LEP\ lower limit is included
by renormalizing the probablity above it, the upper limit is 219 GeV.
The limits show little sensitivity to the inclusion of low $Q^2$ data as
the shift in the prediction is comparable with the theoretical
uncertainty. 
\vspace{-0.3cm}
\vspace{-0.1cm}
\section{Conclusions}
\vspace{-0.2cm}
\Zz\ physics, in the two fermion final state is understood at an
impressive level. The final  results from \LEP\ determine the 
\Zz\ lineshape parameters determined at the sub-per mil level. 
Two fermion physics and W pair production and decay are well
understood above the \Zz\ pole. The measurement of the mass and the width
of the W boson is final for three experiments out of four. The best
direct W mass measurement is achieved  at \LEP\ (Mw = 80.388 $\pm$
0.035 GeV) with a 0.4 per mil relative uncertainty.

The general SM picture shows good global consistency pointing at an
expected Higgs mass below 219 GeV. The standing discrepancies (
$A_{fb}^{0b}$, $W\rightarrow$\taunu\ branching ratio and NuTeV ) show
the need for new data and additional theoretical effort. 
\LEP\ confirms its extremely successful record in
thoroughly testing the standard model by the coherent 
combination of results from its four experiments.
\vspace{-0.5cm}
\section*{Acknowledgments}
\vspace{-0.2cm}
The author would like to acknowledge the support of the European Union grant 
for young scientists. Richard Kellogg, Martin Gr{\"u}newald, Luca
Malgeri, Arno Straessner and  Pat Ward are to be thanked for
enlightening conversations and comments.

\end{document}